\documentstyle[multicol,prb,aps,epsfig]{revtex}

\begin{document}
\draft

\title{Determination of the magnetization scaling exponent for single 
crystal La$_{0.8}$Sr$_{0.2}$MnO$_3$ by broadband microwave surface 
impedance measurements}
\author{Andrew Schwartz,\cite{email} Marc Scheffler,\cite{ms-address} 
and Steven M. Anlage}
\address{Center for Superconductivity Research and MRSEC, 
Department of Physics, University of Maryland\\College Park, MD 20742-4111}
\date{Received 4 June 1999; to be published in Phys. Rev. B Rapid Communications}
\maketitle

\begin{abstract}
Employing a broadband microwave reflection configuration, we have 
measured the complex surface impedance, $Z_S(\omega,T)$, of single 
crystal La$_{0.8}$Sr$_{0.2}$MnO$_3$, as a function of frequency 
(0.045--45~GHz) and temperature (250--325~K). Through the dependence of 
the microwave surface impedance on the magnetic permeability, 
$\hat\mu(\omega,T)$, we have studied the local magnetic behavior of this 
material, and have extracted the spontaneous magnetization, $M_0(T)$, in 
{\em zero applied field}. The broadband nature of these 
measurements and the fact that no external field is applied to 
the material provide a unique opportunity to 
analyze the critical behavior of the spontaneous magnetization at 
temperatures very close to the ferromagnetic phase transition. 
We find a Curie temperature $T_C=305.5\pm 0.5$~K 
and scaling exponent $\beta=0.45\pm 0.05$, in agreement 
with the prediction of mean-field theory. We also discuss other recent
determinations of the magnetization critical exponent in this and
similar materials and show why our results are more definitive.
\end{abstract}


\begin{multicols}{2}
\columnseprule 0pt
\narrowtext

Since the recent discovery of large negative magnetoresistance in the 
manganite perovskites La$_{1-x}$A$_x$MnO$_3$ (where A is typically Ca, 
Sr, or Ba)\cite{vonHelmolt93,Jin94}, much attention has been paid to 
understanding the properties of these materials.\cite{Ramirez97} In 
addition to being potentially useful in technological applications, these 
so-called colossal magnetoresistive (CMR) oxides provide a system in 
which to study electronic and magnetic correlations and the interplay 
between magnetism and transport properties. In particular, a series of 
measurements in recent years have focused on the critical behavior of the 
magnetization in the vicinity of the ferromagnetic phase 
transition.\cite{Martin96,Heffner96,Lofland97a,Lofland97b,Ramos98,Mohan98,Vasiliu-Dolac98,Ghosh98}
These experiments, employing a variety of techniques, have given 
widely varying values of the magnetization scaling exponent $\beta$ 
ranging from about 0.3 to 0.5. This range encompasses both the 
long-range interactions of mean-field theory 
($\beta=0.5$)\cite{Stanley71,Domb96} and the values of 
$\beta$ which result from calculations based on shorter range 
interactions, such as the Ising and Heisenberg models ($\beta=0.325$ and 
0.365, respectively).\cite{LeGuillou80} 

In this paper we present the results of broadband, non-resonant microwave 
surface impedance measurements, in which we have quantitatively 
determined the complex surface impedance, $\hat{Z}_S=R_S+iX_S$, of 
La$_{0.8}$Sr$_{0.2}$MnO$_3$ single crystals over three decades in 
frequency, and as a function of temperature. In contrast to conventional 
magnetization measurements, which typically require the application
of an external magnetic field, this technique allows us to extract
the temperature dependence of the static spontaneous magnetization in 
{\em zero} applied field. We have found that the spontaneous 
magnetization is zero above $T_C$ and rises continuously below $T_C$ in a 
manner which is well described by the theory of critical phenomena near a 
second-order phase transition. From these data we are able to determine 
the scaling exponent $\beta$, and find that it is consistent with the 
value predicted by mean-field theory.

The single crystals of La$_{0.8}$Sr$_{0.2}$MnO$_3$ used in this study 
were grown by the floating-zone technique\cite{Balbashov96} and the 
stoichiometry and structural integrity have been checked by x-ray 
diffraction and energy dispersive x-ray analysis. Disk-shaped samples 
were cut from a 4~mm diameter rod, and resistivity, ac susceptibility, 
and dc magnetization measurements have been reported earlier on samples 
cut from the same boule.\cite{Lofland97b} La$_{0.8}$Sr$_{0.2}$MnO$_3$ has 
a ferromagnetic phase transition with a Curie temperature $T_C$ of 
approximately 305~K. It is well established that the low temperature 
phase is a ferromagnetic metal, while above $T_C$ the system is 
paramagnetic, and the resistivity exhibits a negative slope with respect 
to temperature. The resistivity has a maximum around $T_p=318$~K, 
significantly above $T_C$, which is typical in these manganite 
materials.\cite{Lofland97b} 

In order to determine the temperature and frequency dependence of the 
complex surface impedance we have measured the complex reflection 
coefficient. We have reported the details of the experimental geometry 
elsewhere,\cite{Booth94} and will therefore give only a brief overview of 
the technique here. A phase-locked signal from an HP8510C Vector Network 
Analyzer (45MHz--50GHz) is sent into a coaxial transmission line which is 
terminated by the sample inside a continuous flow cryostat. The amplitude 
and phase of the reflected signal are measured as functions of frequency 
and temperature, and the complex reflection coefficient, 
$\hat{S}_{11}(\omega,T)$, is determined as the ratio of the reflected to 
incident signals. The complex surface impedance of the terminating 
material can be calculated from the reflection coefficient as follows: 
$\hat{Z}_S/Z_0=(1+\hat{S}_{11})/(1-\hat{S}_{11})$, where $Z_0=377\Omega$ 
is the impedance of free space. Due to the phase-sensitive detection 
capabilities of the network analyzer, it is possible to extract {\em 
both} the surface resistance $R_S(\omega)$ and the surface reactance 
$X_S(\omega)$, and the well-defined geometry allows for {\em quantitative} 
evaluation of these material parameters, opening a unique window to 
dynamical processes within the material. 

In the presence of a magnetic field the microwave properties of 
ferromagnetic materials are characterized by two distinct features which 
result from the dispersion of the complex magnetic permeability 
$\hat\mu(\omega)=\mu_1(\omega)-i\mu_2(\omega)$: ferromagnetic resonance 
(FMR) and ferromagnetic anti-resonance (FMAR).\cite{Lax62} At the FMR 
frequency, $\omega_r$, the surface resistance, $R_S(\omega_r)$, shows a 
maximum due to a maximum in the imaginary part of the permeability. At 
the same frequency, the real part of the permeability has a zero-crossing 
with negative slope, as does the surface reactance, $X_S(\omega_r)$. In 
order to satisfy the condition $\mu_1(\omega\rightarrow\infty)=1$, it is 
necessary that there be another zero-crossing, with positive slope, at a 
frequency $\omega_{ar}>\omega_r$. For a ferromagnetic metal, this zero in
$\mu_1$ leads to a {\em reduction} in $R_S$ below the value it would have for
a non-magnetic metal with the same resistivity. This suppression of $R_S$ in
the vicinity of $\omega_{ar}$ is commonly referred to as the ferromagnetic 
{\em anti}-resonance.  Both $\omega_r$ and $\omega_{ar}$ depend not only on the 
externally applied field but also on the local internal magnetization of 
the material, and measurements of the microwave surface impedance 
therefore yield information about the magnetization of material under 
study.

The precise dependence of $\omega_r$ and $\omega_{ar}$ on the 
magnetization $M_0$ can be determined by starting from the 
Landau-Lifshitz-Gilbert equation of motion for the magnetization vector 
in the presence of both a static magnetic field $H_0$ and an oscillatory 
microwave field.\cite{Lax62,Scheffler98,Schwartz99} The complex dynamic 
susceptibility of such a system can be written in the following form:\cite{Lax62}
\begin{equation}
\hat\chi(\omega)=\frac{\hat\mu(\omega)}{\mu_0}-1 = 
\frac{\omega_M[(\omega_0+i\Gamma)+\omega_M]} 
{\omega_r^2-\omega^2+i\Gamma[2\omega_0 + \omega_M ] },
\label{eq:Chi}
\end{equation}
where $\omega_M=\gamma \mu_0 M_0$, $\omega_0=\gamma \mu_0 H_0$, 
$\omega_r=\sqrt{\omega_0(\omega_0+\omega_M)}$, $\Gamma=\alpha\omega$, 
$\alpha$ is the dimensionless Gilbert damping parameter, and $\gamma$ is 
the gyromagnetic ratio for an electron. We have expressed the field and
magnetization as frequencies in order to clarify the comparison to our 
frequency-dependent data. It is clear from Eq.~(\ref{eq:Chi}) that for 
small damping the quantity $\omega_r$ is the ferromagnetic resonance 
frequency, and it can be shown that the anti-resonance frequency, 
the point at which $\mu_1=0$, is given by $\omega_{ar}=\omega_0+\omega_M$. 
For simplicity, Eq.~(\ref{eq:Chi}) has been written for the limiting case of an 
infinitely thin sample with the static magnetic field applied in the 
plane of the sample. For a finite sized sample there are corrections to 
this form due to demagnetization effects.\cite{Lax62}

As discussed above, the microwave reflection measurement which we have 
employed yields the surface impedance instead of the permeability, 
however the two are related as follows: 
$\hat{Z}_S(\omega,T)=\sqrt{i\omega\hat{\mu}(\omega,T)\rho}$. We have 
assumed that at microwave frequencies La$_{0.8}$Sr$_{0.2}$MnO$_3$ is in 
the Hagen-Rubens limit (i.e. $\rho_2\ll\rho_1 \approx \rho_{\rm dc}$), 
allowing us to insert a frequency-independent value for $\rho$. Then we 
can substitute the expression for the susceptibility from 
Eq.~(\ref{eq:Chi}) into this expression in order to model the complete 
frequency dependence of $R_S$ and $X_S$. Measurements of the surface 
impedance as a function of applied magnetic field have shown that this 
model provides an excellent description of the measured 
data.\cite{Scheffler98,Schwartz99}

\begin{figure}[htb]
\begin{center}
\leavevmode
\epsfig{file=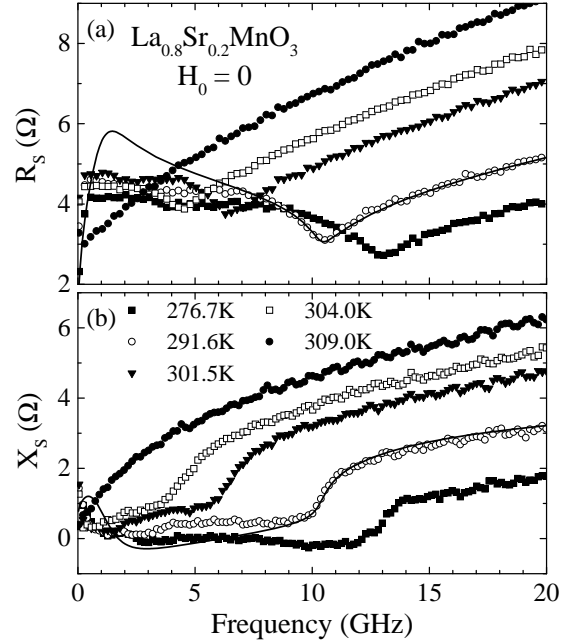,width=7.5cm,clip=,bbllx=100pt,bblly=140pt,
bburx=535pt,bbury=640pt}
\end{center}
\caption{\label{fig:RsXs_f}
The zero-field frequency dependence of (a) the surface 
resistance $R_S$ and (b) the surface reactance $X_S$ at various 
representative temperatures. The solid lines show a fit to the data
at 291.6~K, as discussed in the text.}
\end{figure}

Figure~\ref{fig:RsXs_f} shows zero-field $R_S(\omega)$ and $X_S(\omega)$ 
spectra at various temperatures. The anti-resonance features, minima in 
$R_S$ and steps in $X_S$ are clearly visible at all temperatures below 
$T_C$, despite the fact that $H_0=0$. Naively we would expect the 
permeability to be dispersionless in the absence of a static magnetic 
field, and thus both $R_S(\omega)$ and $X_S(\omega)$ should have the 
square root frequency dependence characteristic of a metal, as is seen in 
the 309~K spectra, above $T_C$. We can understand why this is not so if 
we consider that even in the absence of an {\em external} magnetic field, 
there are local internal fields $H_i$ due to anisotropy and domain 
structure, for example. Although 180-degree domain walls separating 
magnetically saturated regions can in principle produce rather large 
internal fields,\cite{Lax62} our field-dependent measurements of 
the anti-resonance as $H_0\rightarrow 0$ allow us to estimate that 
$\mu_0 H_i\leq 0.02$~T,\cite{Schwartz99} probably due to a more
complicated domain structure. Such small but finite fields 
can, however, cause the precession of the magnetization and 
thereby the dispersion of the permeability. 
Since $\omega_{ar}=\gamma\mu_0(H_i+M_0)$ and $H_i\ll M_0$, we see
a well-defined anti-resonance feature at a frequency determined 
predominantly by $M_0$ and a width due to inhomogeneities in the
internal fields and the intrinsic damping given by $\alpha$.  
Therefore we can extract the temperature dependence of the magnetization,
in the absense of an applied field, from the temperature dependence 
of the anti-resonance frequency.

\begin{figure}[htb]
\begin{center}
\leavevmode
\epsfig{file=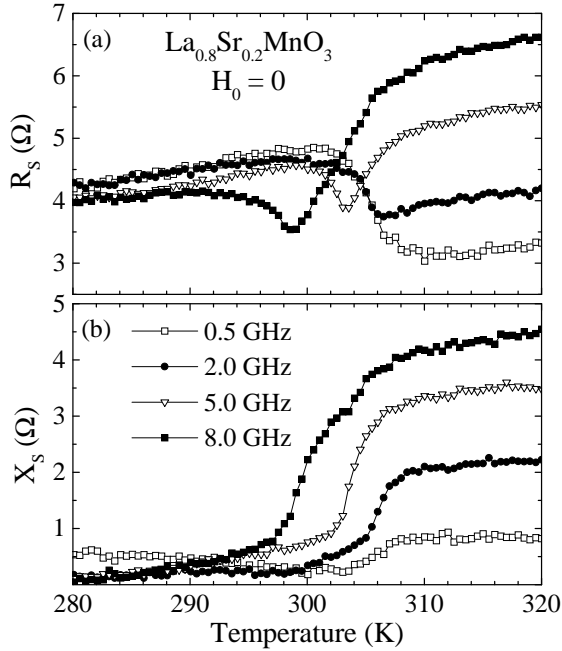,width=7.5cm,clip=,bbllx=100pt,bblly=140pt,
bburx=535pt,bbury=645pt}
\end{center}
\caption{The zero-field temperature dependence of (a) the surface 
resistance $R_S$ and (b) the surface reactance $X_S$ at various 
representative frequencies.
\label{fig:RsXs_T}}
\end{figure}

The solid lines on Fig.~\ref{fig:RsXs_f} are fits of the model presented
above to the 291.6~K spectra, where we have set $H_0=0$, but included a 
finite damping to account for a distribution of internal fields around 
$H_i=0$. The discrepency between the model and the data at low frequencies
is probably due to the fact that the model does not properly describe this
distrubution, but it is clear that the model does an excellent job of describing 
the behavior of both components of the surface impedance in the vicinity 
of the ferromagnetic anti-resonance. 

The surface impedance can also be measured as a function of temperature 
at fixed frequency, and Fig.~\ref{fig:RsXs_T} shows such data at a few 
representative frequencies. The anti-resonance is manifested as a minimum 
in $R_S(T)$ coincident with an inflection point in $X_S(T)$, and moves to 
lower temperature as the frequency increases. It is, of course, also 
possible to extract $M_0(T)$ from these data, and therefore we have
four sets of spectra from which to determine $M_0(T)$. 

\begin{figure}[htb]
\begin{center}
\leavevmode
\epsfig{file=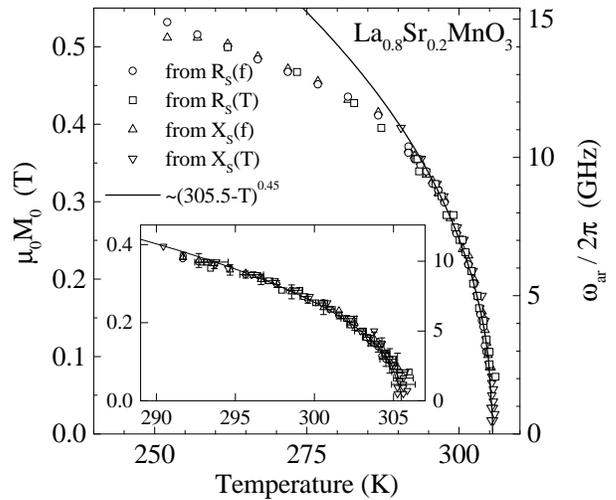,width=8cm,clip=,bbllx=155pt,bblly=70pt,
bburx=665pt,bbury=500pt}
\end{center}
\caption{The spontaneous magnetization $M_0$ of 
La$_{0.8}$\-Sr$_{0.2}$\-MnO$_3$ extracted from the four sets of data shown in 
Figs.~\protect\ref{fig:RsXs_f} and \protect\ref{fig:RsXs_T} (for each 
data set there are many more spectra than are shown in these figures). 
The inset shows the same data over a smaller temperature range near 
$T_C$, and the solid lines are a fit to the data with the scaling form 
given in the text. The right axes show the FMAR frequencies. 
\label{fig:M_of_T}}
\end{figure}

For the $X_S$ data, both as a function of frequency and temperature, we 
have determined $\omega_{ar}$ by finding the peak of the first 
derivative, $dX_S/df$ [Fig.~\ref{fig:RsXs_f}b] or $dX_S/dT$ 
[Fig.~\ref{fig:RsXs_T}b]. Similarly, $\omega_{ar}$ is determined from the 
positions of the local minima in $R_S(f)$ and $R_S(T)$. 
Figure~\ref{fig:M_of_T} shows the magnetization curve which we have 
extracted from these four sets of data. The onset of spontaneous 
magnetization at $T_C\approx 305.5$~K is clearly seen.

With these data it is possible to examine the critical 
behavior of the magnetization in the vicinity of $T_C$. The theory of 
critical phenomena at a second order phase transition predicts that the 
magnetization will vary as a power law, $M_0(T) \propto (T_C-T)^\beta$, 
where $\beta$ depends on the Hamiltonian describing the interactions 
among the spins.\cite{Stanley71,Domb96} Therefore, determination of this 
exponent yields information about the range of ferromagnetic 
interactions. But this expression is expected to hold only in the limit 
$T \rightarrow T_C^-$, so it is necessary to look at the behavior very 
close to $T_C$, where the slope of $M_0(T)$ is very large. One approach 
is to calculate the following function:\cite{Kouvel64} 
$T^*(T)=-M_0(dM_0/dT)^{-1}=\beta^{-1}(T_C-T)$. Thus $T^*$ is linear in 
$T$, and values of $\beta$ and $T_C$ can be determined 
from the slope and intercept. For the data shown in Fig.~\ref{fig:M_of_T} 
we find that $T^*$ is roughly linear for temperatures above 290K, and 
below this temperature the slope of $T^*(T)$ increases due to the 
saturation of the magnetization. It is therefore only reasonable to 
examine the critical behavior in the temperature region between 290K and 
$T_C$. If we define the dimensionless quantity $\varepsilon=1-T/T_C$ then 
we see that this corresponds to a range of $\varepsilon\approx 0-0.05$. 
By fitting a straight line to this portion of the $T^*(T)$ curve we find 
that $T_C=305.5\pm 0.5$~K and $\beta=0.45\pm 0.05$. The solid line on 
Fig.~\ref{fig:M_of_T} shows the critical behavior with these parameters, 
and the inset shows the same data and model between 290K and $T_C$, with 
some representative error bars included.

A previous study of the microwave properties of similar 
La$_{0.8}$Sr$_{0.2}$MnO$_3$ single crystals gave values of $T_C=304\pm 
3$~K and $\beta=0.34\pm 0.05$.\cite{Lofland97b} This value of $\beta$ is 
clearly in disagreement with ours, even though the $M_0(T)$ data from the 
two experiments are in agreement. This apparent contradiction can be 
understood as follows: in the earlier study, the value of $\beta$ was 
obtained by fitting the magnetization data between 270 and 300K 
($\varepsilon\approx 0.01-0.11$). As we have shown above, the use of data 
below 290K is not appropriate for an examination of the critical 
behavior. In addition, our ability to measure FMAR to lower frequencies 
allows us to determine the magnetization much closer to $T_C$, which is 
the most important region for an accurate determination of both $\beta$ 
and $T_C$. If we fit our data only over the range $\varepsilon\approx 
0.01-0.11$, then we find $T_C=303$~K and $\beta=0.26$. If instead we fix 
$T_C$ at 305.5~K and allow only $\beta$ to vary, again fitting over the 
same range, we find $\beta=0.33$. So it is clear that the difference 
between our result and that of the previous measurement arises simply from 
the temperature range over which the analysis of the critical behavior 
has been done. 

Our experimental value of the scaling exponent $\beta$ is in agreement 
with mean-field theory, which predicts 
$\beta=0.5$.\cite{Stanley71,Domb96} Similar results were obtained from 
microwave measurements on La$_{0.7}$Sr$_{0.3}$MnO$_3$ single crystals, 
which gave $\beta=0.45\pm 0.05$.\cite{Lofland97a} Their fit range was 
$\varepsilon\approx 0.01-0.12$. And recently reported dc magnetization 
measurements on polycrystalline La$_{0.8}$Sr$_{0.2}$MnO$_3$, also gave 
$\beta=0.50\pm 0.02$, using a narrow temperature range of 
$\varepsilon\approx 0-0.01$.\cite{Mohan98} However, neutron scattering 
and dc magnetization measurements on La$_{0.7}$Sr$_{0.3}$MnO$_3$ single 
crystals yielded $\beta=0.295\pm 0.002$ ($\varepsilon\approx 0-0.13$) and 
$\beta=0.37\pm 0.04$ ($\varepsilon\approx 0-0.03$), 
respectively.\cite{Martin96,Ghosh98} Recent neutron scattering 
measurements on single crystals of both La$_{0.8}$Sr$_{0.2}$MnO$_3$ and 
La$_{0.7}$Sr$_{0.3}$MnO$_3$ gave values of $\beta=0.29\pm 0.01$ 
($\varepsilon\approx 0-0.18$) and $\beta=0.30\pm 0.02$ 
($\varepsilon\approx 0-0.3$), respectively.\cite{Vasiliu-Dolac98} 
Finally, a measurement of the temperature dependence of the zero-field 
muon precession frequency in La$_{0.67}$Ca$_{0.33}$MnO$_3$ gave 
$\beta=0.345\pm 0.015$ ($\varepsilon\approx 
0.03-0.27$).\cite{Heffner96} Although many of these studies yielded values 
of $\beta$ which are significantly lower than what we have found, the 
explanation for this seems to lie in the fact that the saturation of the 
magnetization will always lead to a reduced value for $\beta$ if the data 
are fit over too wide a temperature range.

In conclusion, we have extracted the zero-field spontaneous magnetization 
of single crystal La$_{0.8}$Sr$_{0.2}$MnO$_3$ from the temperature and 
frequency dependence of the microwave surface impedance. This 
magnetization rises continuously below $T_C$, as expected for a 
second-order phase transition. Analysis of these data gives values for the Curie 
temperature $T_C=305.5\pm 0.5$~K and the scaling exponent $\beta=0.45\pm 
0.05$. Unfortunately, it seems that there is not yet experimental 
consensus about the scaling behavior of the magnetization in these 
compounds. It is clear that the value of $\beta$ is very sensitive to the 
fit range, with values ranging from $\beta=0.3-0.5$ reported in the 
literature. The technique presented here is unique because it requires
no external field and is broadband, both of which allow us to examine the
asymptotic critical behavior of the magnetization as $T\rightarrow T_C$.
The result presented here is consistent with the mean-field 
value (0.5), implying that there are long-range 
ferromagnetic interactions in this lanthanum manganite.

We thank Y. Mukovskii and colleagues for growing the 
samples used in this study. We also thank S. Bhagat and S. Lofland 
for providing the samples to us, and for many useful discussions, 
and R. Greene and C. Lobb for their helpful comments. 
This work was supported by NSF DMR-9624021, the Maryland/NSF MRSEC
(NSF DMR-9632521), and the Maryland Center for Superconductivity Research.

\end{multicols}
\end{document}